\documentclass[doublespacing]{elsart}
\usepackage{natbib}
\usepackage{amssymb}
\usepackage{amsfonts}
\usepackage{graphicx}

\begin{document}

\begin{frontmatter}


\title{Eutacticity in sea urchin evolution}


\author{J. L\'opez-Sauceda, J.L. Arag\'on\corauthref{cor1}}
\corauth[cor1]{Author for correspondence.}
\address{Departamento de Nanotecnolog\'{\i}a, \\
  Centro de F\'{\i}sica Aplicada y Tecnolog\'{\i}a Avanzada, \\
  Universidad Nacional Aut\'onoma de M\'exico, \\
  Apartado Postal 1-1010, Quer\'etaro 76000, M\'exico.}

\begin{abstract}
  An eutactic star, in a $n$-dimensional space, is a set of $N$
  vectors which can be viewed as the projection of $N$ orthogonal
  vectors in a $N$-dimensional space. By adequately associating a star
  of vectors to a particular sea urchin we propose that a measure of
  the eutacticity of the star constitutes a measure of the regularity
  of the sea urchin. Then we study changes of regularity (eutacticity)
  in a macroevolutive and taxonomic level of sea urchins belonging to
  the Echinoidea Class. An analysis considering changes through
  geological time suggests a high degree of regularity in the shape of
  these organisms through their evolution. Rare deviations from
  regularity measured in Holasteroida order are discussed.
\end{abstract}

\begin{keyword}
Eutactic stars \sep Bilateral symmetry \sep Regularity \sep Sea urchins
\end{keyword}
\end{frontmatter}


\section{INTRODUCTION}
\label{sec:intro1}

This work is dealing with regularity, which is a property with deep
implications in organisms. From the biological point of view
regularity has been related with radial symmetry, and irregularity
with bilateral symmetry \citep{Coen}. The heuristic value of radial
and bilateral symmetry in biology account for taxonomic issues,
however, symmetry as well as disruption symmetry have been an
empirical and intuitive approach accounting for structural properties
in organisms \citep{Holland,Smith,Jan,Rasskin,Knoblich}.

From a mathematical point of view, the property of regularity of a
geometric form has not been formalized. Based in previous results by
\citet{Torres}, we hypothesize that \emph{eutacticity} provides a
measure of regularity based in the following argument. A set of $N$
vectors in $\mathbb{R}^n$, with a common origin, is called a star and
a star is said to be eutactic if it can be viewed as the projection of
$N$ orthogonal vectors in $\mathbb{R}^N $. It turns out that stars
associated with regular polygons, polyhedra or, in general, polytopes,
are eutactic \citep{Coxeter} and thus regularity and eutacticity are
closely linked. A disadvantage of using eutacticity as a measure of
regularity is that a star vector must be associated with the
geometrical form under study. As we shall see, this is not a problem
with echinoids. In fact, \citet{Torres} found that the flower-like
patterns formed by the five ambulacral petals in 104 specimens of
plane irregular echinoids (from Clypeasteroidea) are eutactic. Here we
present a deeper study that overcome the restriction to plane
irregular echinoids, using the five ocular plates (OP) to define the
star vector. Additionally, we use a new criterion of eutacticity that
provides a measure of the degree of eutacticity of a star which is not
strictly eutactic. With these tools we study the variability of
eutacticity during geological time and to analyze pentamery
variability during the evolution of sea urchins.

Sea urchins are pentameric organisms with an apical structure, called
the apical disc \citep{Melville}. This structure includes five ocular
plates (OP) that can fold the vector star associate with each sea
urchin species (see Fig. \ref{fig:fig1} and Section \ref{sec:discoap}
for a detailed description). In this work, we show that OP can be
useful even in ovoid echinoids, such as Spatangoids, since the OP are
almost tangential to the aboral surface (opposite to oral
surface). Using the OP to define the star of vectors, we analyze the
regularity and changes in a macroevolutive and taxonomic level in a
collection of 157 extinct and extant sea urchins. We conclude that
evolution has preserved a high degree of regularity and, consequently,
that the apical disk is a homogeneous and geometrically stable
structure through the geological time. Low values of regularity were
recorded in some specific families and its biological consequences are
discussed.

This paper is organized as follows. In Section \ref{sec:eutactic} a
mathematical introduction to the concept of eutactic star is
presented.  Section \ref{sec:discoap} describes the structure of the
apical disc and its biological importance, making it the obvious
choice to define a vector star which characterizes each
specimen. Experimental methods and results are devoted to Section
\ref{sec:resultados} and, finally, discussion and conclusions are
presented in Section \ref{sec:discusion}.

\section{REGULARITY AND EUTACTIC STARS}
\label{sec:eutactic}

Our main hypothesis is that the concept of regularity of a biological
form may play an important role in the study of phenotipic variation in
evolution. For this goal, one must first be able to establish a formal
criterion defining regularity of a geometrical form, including a
measure of how regular a form is. Mathematically, this property has
not been defined and here, as a first step along this direction, we
adopt the concept of eutacticity that, as we shall show, is closely
related to regularity.

We shall deal with a set of $N$ vectors $\left\{ \mathbf{a}_1,
  \mathbf{a}_2, \ldots, \mathbf{a}_N \right\}$ in $\mathbb{R}^n$, with
a common origin, called \emph{star}. In this case $N>n$ so the set of
vector con not be linearly independent. The star is called
\emph{eutactic} if its vectors are orthogonal projections of $N$
orthogonal vectors in $\mathbb{R}^N$, that is, there exist $N$
orthogonal vectors $\left\{ \mathbf{u}_1, \mathbf{u}_2, \ldots,
  \mathbf{u}_N \right\}$, in $\mathbb{R}^N$, and an orthogonal
projector $P: \mathbb{R}^N \rightarrow \mathbb{R}^n$ such that
\[
P(\mathbf{u}_i) = \mathbf{a}_i, \;\;\;\; i = 1,2, \ldots, N. 
\]

The notion of eutacticity (from the Greek \emph{eu}=good and
\emph{taxy}=arrangement) was firstly introduced by the Swiss
mathematician L.  Schl\"afli (about 1858) in the context of regular
polytopes. Later, \citet{Hadwiger} noticed that the vectors of an
eutactic star are projections from an orthogonal basis in higher
dimensional spaces and proved that the star associated to a regular
polytope is eutactic. Thus, eutacticity is associated with regularity
and the remarkable properties of eutactic stars have been useful in
different realms such as quantum mechanics, sphere packings,
quasicrystals, graph and frame theory and crystal faceting (see
\citet{Aragon1} and references therein).

A well known necessary and sufficient condition for a star to be
eutactic is due to Hadwiger himself, who proved that a star $\left\{
  \mathbf{a}_1, \mathbf{a}_2, \ldots \mathbf{a}_N \right\}$ in
$\mathbb{R}^n$ is eutactic if and only if there is a real number
$\lambda$ such that
\[
\sum _{i=1}^N \left( \mathbf{x} \cdot \mathbf{a}_i \right)
\mathbf{a}_i = \lambda \mathbf{x} ,
\]
is fulfilled for all $\mathbf{x} \in \mathbb{R}^n$. In the special
case where $\lambda = 1$, the star is said to be \emph{normalized
  eutactic}.

A more practical form of the eutacticity criterion is obtained if the
so called structure matrix $A$ is introduced. Let $A$ be the matrix
whose columns are the components of the vectors $\left\{ \mathbf{a}_1,
  \mathbf{a}_2, \ldots, \mathbf{a}_N \right\}$, with respect to a
given fixed orthonormal basis of $\mathbb{R}^n$. In this case, the
matrix form of Hadwiger's theorem sates that the star represented by
$A$ is eutactic if and only if
\begin{equation}  \label{eqn:aat}
A A ^T = \lambda I,
\end{equation}
for some scalar $\lambda$ (here $I$ is the $n \times n$ unit matrix).

In this work we are dealing with stars measured in digital images of
sea urchins and thus a reliable numerical criterion of eutacticity,
suitable to work with experimental measurements, is need. Notice that
a criterion such as (\ref{eqn:aat}) is not useful since experimental
errors may produce a matrix which is not exactly the identity matrix
$I$. Thus, it is desirable to obtain a numerical criterion capable of
measuring the degree of eutacticity of a star which is not strictly
eutactic. This criterion has already been proposed \citep{Aragon1} and
asserts that a star in $\mathbb{R}^n $, represented by the structure
matrix $A$, is eutactic if and only if
\begin{equation}
 \label{eqn:cosfi}
\varepsilon = \frac{\textrm{Tr} (S)}{\sqrt{\textrm{Tr} (SS)} \sqrt{n}} = 1,
\end{equation}
where $S = A A^T$. Notice that the closer $\varepsilon$ is to one, the
more eutactic the star is. In the particular case of two-dimensional
stars ($n=2$), it can be proved that $1/\sqrt{2} \leq \varepsilon \leq
1$ \citep{Aragon1}.

\section{SEA URCHINS AND VECTOR STARS}

\label{sec:discoap}

In \citet{Torres}, vector stars were associated to the petaloid
ambulacra of plane irregular echinoids. It was reported that, for 104
specimens of the Natural History Museum of London, the pentagonal
stars thus defined fulfill very accurately an eutacticity
criterion. The calculations carried out in that work present two main
restrictions: a) stars are associated to plane or almost plane sea
urchin specimen and b) A eutacticity criterion was used that depends
on the coordinate system and does not allow a measurement of the
degree of eutacticity of stars which are not strictly eutactic. Here
we overcome these restrictions by using the eutacticity criterion of
(\ref{eqn:cosfi}) and using the five ocular plates (OP) to associate a
star of vectors to each sea urchin. As we shall see in what follows,
besides the biological importance of the OP, its use to define a star
of vectors allows to study non planar echinoids.

\begin{figure}[t!]
\begin{center}
  \includegraphics[width=10.0cm]{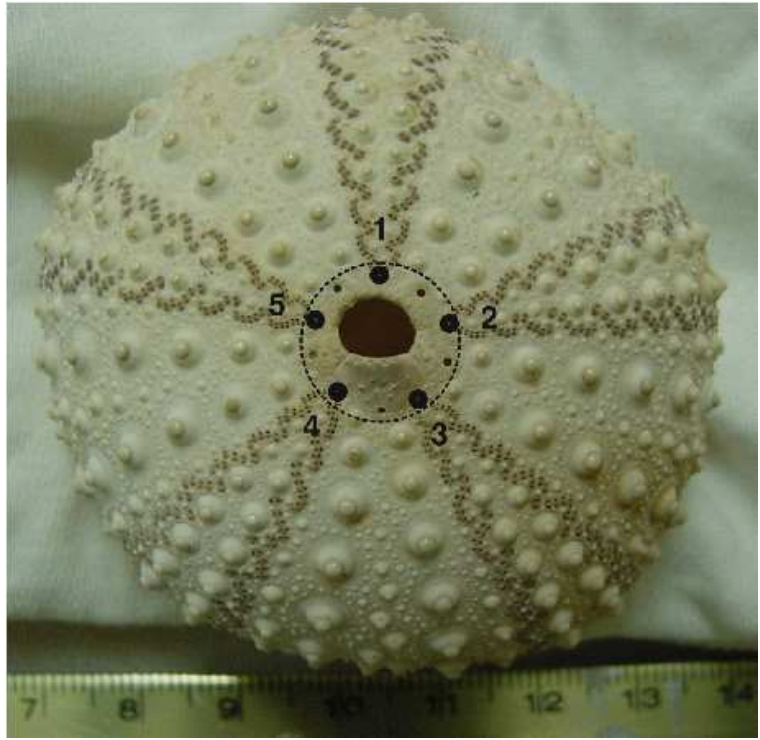}
\end{center}
\caption{Apical disc (encircled) showing landmarks (numbered) at ocular
  plates. The star vector is formed by vectors pointing to these
  landmarks, with common origin at the center of mass.}
\label{fig:fig1}
\end{figure}

The apical disc in sea urchins, encircled in Fig. \ref{fig:fig1},
represents a crown of biological structures in the apex of the test
\citep{Melville}. It is positioned in the aboral surface of the test
and is conformed by five genital plates, five ocular plates and the
madreporite. The OP are located at the point of origin of the
ambulacral zones. The biological relevance of ambulacra is given by
the following reasons. Firstly, each ambulacra consist of two, or even
more, columns of plates extending from the margin of an OP to the edge
of the mouth. In most echinoids each mature plate is perforated by two
pores forming a pore pair; each pore pair gives passage to one tube
foot, which is connected internally with the water vascular
system. Secondly, five ambulacra are a conspicuous sign of body plan
pentamery in all extant and extinct sea urchins and, finally,
ambulacral rays are seen as homologous structures in echinoderms. Thus
we conclude, that OP has more biological implications as the origin of
ambulacra than the end of petaloid ambulacra.  In addition, the OP are
almost tangential to the aboral surface and thus it is useful even for
ovoid echinoids. 

We then define the star vector associated to a particular echinoid as
the set of five vectors pointing to the OP with origin at the
centroid. The star thus defined, allows us to test eutacticity in a
wide range of echinoids and study changes in a macroevolutive and
taxonomic level.


\begin{table}[t!]
  \caption{List of the 47 families studied in this work. The number of specimens
    considered in this work is indicated between parenthesis.}
  \label{table:table1}
\begin{tabular}{|llll|}
  \hline
  Radial families: &  & & \\ \hline
  Archaeocidaridae (1) & Aspididisdematidae (1)  & Diadematidae (2) &  Echinometridae (2)\\ 
  Lissodiadematidae (1) & Micropygidae (1) & Parechinidae (1) & Psychocidaridae (1) \\ 
  Saleniidae (1) & Toxopneustidae (3)  & Toxopneustidae (3) & Temnopleuridae (2)  \\ \hline
  Bilateral families: &  & & \\ \hline
  Arachnoidae (5) & Archiaciidae (3) & Asterostomatida (1) & Astriclypeidae (2) \\
  Brissidae (8) & Cassiduloidae (12) & Clypeasteridae (12) & Clypeolampaidae (3) \\
  Collyritidae (4) & Corystidae (1)  & Dendrasteridae (2) & Disasteridae (3)  \\
  Echinarachniidae (4) & Eoscutellidae (1) & Fibulariidae (4) & Galeropygidae (4) \\
  Hemiasteridae (7) & Holasteridae (2) & Holectypidae (4) & Laganidae (3) \\
  Loveniidae (1) & Mellitidae (1) & Micrasteridae (6) & Neolampadidae (2) \\
  Neolaganidae (5) & Nucleolitidae (3) & Pliolampadidae (2) & Pourtalesiidae (2) \\
  Pygasteridae (5) & Protoscutellidae (2) & Rotulidae (3) & Scutellidae (4) \\
  Schizasteridae (9) & Somaliasteridae (1) & Spatangoidae (2) & Toxasteridae (8) \\ \hline 
\end{tabular}
\end{table}

\section{RESULTS}

\label{sec:resultados}

\subsection{Variability in taxonomic groups}

We have analyzed 157 extant and extinct specimens of sea urchins from
the collection of the Instituto de Ciencias del Mar (Universidad
Nacional Aut\'{o}nma de M\'{e}xico) and from images of The Natural
History Museum of London web site (http://www.nhm.ac.uk). As shown in
Table \ref{table:table1}, the analyzed sea urchins belong to 47
Families in a taxonomic group of 95 Families, according to the
classification by \citet{Lebrun}. Eleven of these specimens are radial
and thirty six bilateral. To each sea urchin we associate a vector
star, with vectors pointing to the OP (shown in Fig. \ref{fig:fig1})
and origin at the centroid. Measurements were carried out on digital
images of aboral surfaces, analyzed using the morphometric software
packages \emph{MakeFan6} \citep{Zelditch} and \emph{tpsDig2}
\citep{Rohlf}. With the former, the OP is digitized and the vector
star coordinates are obtained by using the second program. Once the
coordinates of the star vector are available, Eq. (\ref{eqn:cosfi}) is
used to calculate the value of eutacticity of the star, \emph{i.e.},
$\varepsilon $.

\begin{figure}[t!]
\begin{center}
\includegraphics[width=10.0cm]{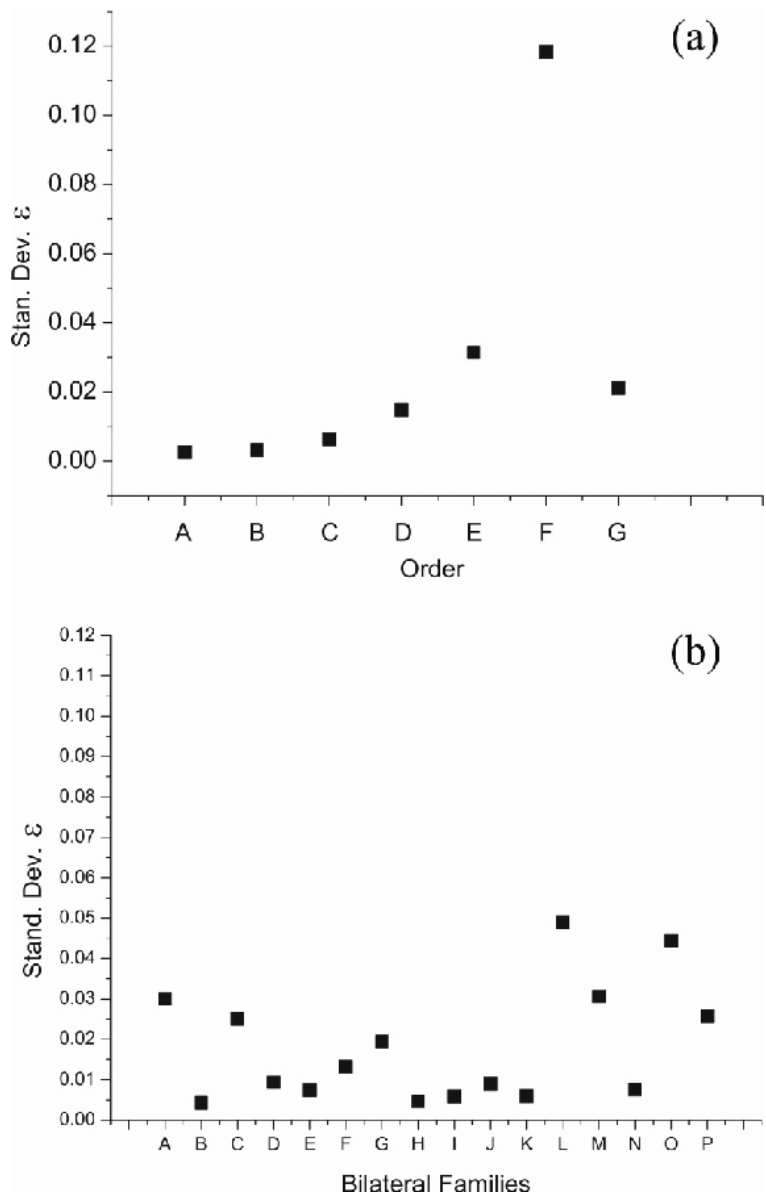}
\end{center}
\caption{(a) Standard deviation of $\varepsilon$ per oder: A)
  Clypeastroida, B) Cassiduoida, C) Holectypoida, D) Pygasteroida, e)
  Spatangoida, F) Holasteroida, G) Disasteroida. (b) Standar deviation
  of $\varepsilon$ per family (the bilateral families with the largest
  variations are only displayed.): A) Pygasteridae, B) Holectypidae,
  C) Cassiduloidae, D) Clypeasteridae, E) Arachnoidae, F)
  Fibulariidae, G) Neolaganidae, H) Rotulidae, I) Echinarachniidae, J)
  Scutellidae, K) Collyritidae, L) Toxasteridae, M) Micrasteridae, N)
  Brissidae, O) Hemiasteridae, P) Schizasteridae.}
\label{fig:fig2}
\end{figure}

Since one of our goals is to analize regularity of sea urchins through
geological time, we must define a taxonomic group to define a
phylogenetic reference in time. Taxonomic keys use the apical disk as
a reference to describe the family level, thus family taxonomic level
constitutes the best choice since it should have a low variability in
the apical disk. In fact, as shown in Fig. \ref{fig:fig2}, family
level shows lower variability in eutactic values as compared to
order. Hence, this taxonomic level was used to represent regularity in
geological time.

By organizing the values of eutacticity per family, we are able to
carry out a formal statistical analysis. From the properties of
eutacticity, we can deduce that radial families have a high degree of
eutacticity (the stars form regular pentagons) and, consequently, no
variability (up to experimental errors). Contrarily, high variability
is expected in bilateral families.

Before proceeding with a statistical analysis of the eutacticity
values per family, we have to take into account the possibility of a
stochastic nature of eutacticity. In order to reject this possibility,
an experimental set of two hundred randomly generated bilateral stars
was considered. A formal statistical analysis must then include three
groups: radial, bilateral and random stars. The values of $\varepsilon
$ of the random sample yield a mean of 0.891026 with population
standard error of 0.00604. Our experimental sample of radial stars
yield a mean of 0.995187423 and standard deviation of
0.00526708. Finally, the experimental sample of bilateral stars gives
a mean of 0.96499158 and standard deviation of 0.062079301. The
Shapiro-Wilk test applied to our samples yields $W=0.534789$ and $p
\langle 0.00001$, for the joined radial and bilateral experimental
sample, and $W=0.853001$ and $p\langle 0.00001$, for the random
sample. From this result we conclude that neither the experimental or
random distribution are normal and thus a non parametric statistical
analysis is needed. This non parametric test produces
$\chi_{0.05}^{2}=68.2774$ and $p\langle 0.0001$, consequently, the
probability of finding random significant differences between radial,
bilateral and random stars is lower that 0.0001.  The possibility of a
stochastic origin of regularity is thus rejected.

\begin{figure}[t!]
\begin{center}
\includegraphics[width=10.0cm]{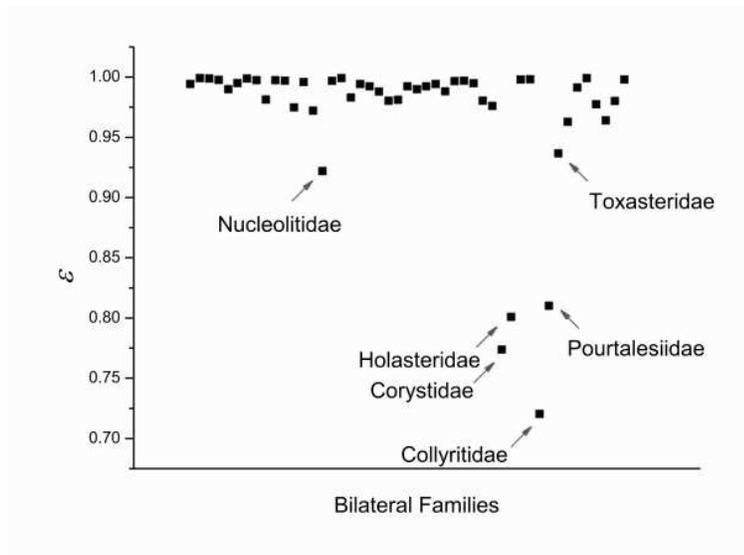}
\end{center}
\caption{Scatter plot of eutactic values in 47 Families of radial and
  bilateral sea urchins. The lowest degree of regularity are recorded
  in Holasteridae, Corystidae, Collyritidae, Pourtoalesiidae,
  Toxasteridae and Nucleolitidae.}
\label{fig:fig3}
\end{figure}

Now, concerning the analysis of the experimental sample per family, in
Fig. \ref{fig:fig3}, a scatter plot of the eutacticity values of the
47 families is shown. From the figure, it is observed that the lowest
degree of regularity are recorded in Holasteridae, Corystidae,
Collyritidae, Pourtoalesiidae, Toxasteridae and Nucleolitidae. This
observation will be revisited in the next Section. A
Wilcoxon/Kruskal-Wallis analysis of the 47 families was carried out,
yielding $\chi_{0.05}^{2}=78.8904$ and $p\langle
0.0001$. Consequently, differences between families are also accepted.

\subsection{Eutacticity values through geological time}

\begin{figure}[t!]
\begin{center}
\includegraphics[width=10.0cm]{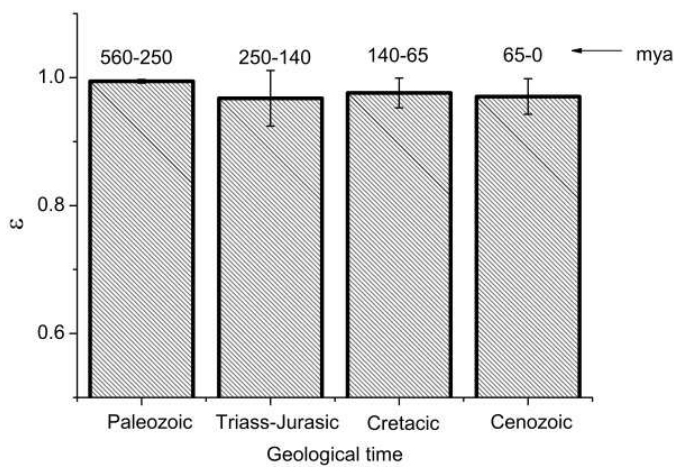}
\end{center}
\caption{Mean values of the eutacticity parameter ($\varepsilon$) of
  the experimental sample in four intervals of the geological time.}
\label{fig:fig4}
\end{figure}

Here the variability of eutacticity per family through geological time
is studied. The experimental sample includes extant and extinct
specimens. Once again we have to take into account that radial
echinoids are associated with nearly eutactic stars; the most
primitive groups, like Paleozoic groups, are almost always totally
radial. Contrarily, the eutacticity values of species from
post-paleozoic groups are less uniform and thus more than two
specimens per family are required. Fig. \ref{fig:fig4} shows the mean
values of $\varepsilon$ at four geological time intervals, namely
Paleozoic, Triasic-Jurasic, Cretaceous and Cenozoic. These scales were
chosen because, according the paleontological records \citep{Lebrun},
at the beginning of each of these intervals there was a rise in the
speciation rate; there was an increase in the numbers of families. As
shown in Fig. \ref{fig:fig4}, post-paleozoic sea urchins show the
highest degree of variability. A statistical analysis, however, gives
the values $\chi_{0.05}^{2}=6.1418$ and $p\langle 0.1049$, implying
that there are no statistical differences in regularity through
geological time.

\begin{figure}[t!]
\begin{center}
\includegraphics[width=10.0cm]{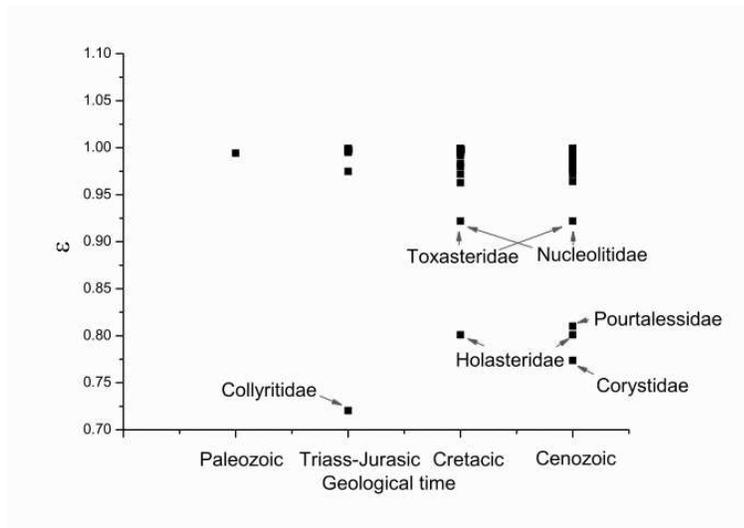}
\end{center}
\caption{Plot of the eutacticity values ($\varepsilon$) per family
  through the geological time.}
\label{fig:fig5}
\end{figure}

\begin{figure}[t!]
\begin{center}
\includegraphics[width=10.0cm]{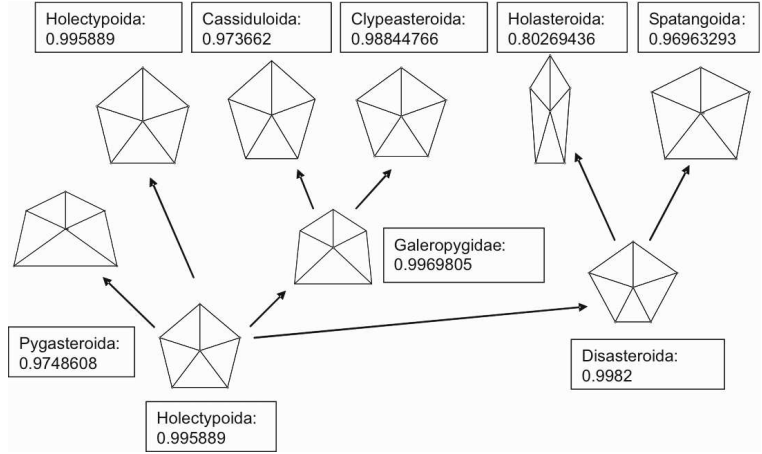}
\end{center}
\caption{Evolutive cladogram of the class Echinoidea, depicted for
  illustrative purposes. A representative star of vectors and the mean
  value of eutacticity per order is included}
\label{fig:fig6}
\end{figure}

In Fig. \ref{fig:fig5} a plot of the eutacticity values, per family,
through the geological time is shown. The lowest values of eutacticity
are recorder firstly in early Mesozoic Collyritidae and low values
continue with late Mesozoic and Cenozoic Holasteridae, Portuolesidae
and Corystidae. As a matter of fact, all these families belong to the
Holasteroida order which turns out to be the responsible of the
prominent peak (F) in the standard deviation plot in
Fig. \ref{fig:fig2}(a). In order to have a better understanding of the
singularity of Holasteroida order, in Fig. \ref{fig:fig6} an evolutive
cladogram showing phylogenetic relationships between orders is
shown. In this cladogram a representative star, and the mean value of
eutacticity per order, is included. It is clearly shown that the
lowest values of regularity comes from Disasteroida and are recorded
in Spatangoida and Holasteroida. In fact, from the measured values, we
can say that Holasteroida is an \textquotedblleft
anti-eutactic\textquotedblright group, \emph{i.e.}, mathematically
irregular. Most living representatives of Holasteroida are deep-water
inhabitants with exceedingly thin and fragile tests. Besides that we
consider that regularity and irregularity constitute two important
parameters to approach ecological and evolutive topics, the observed
departure from regularity could have been a way to increase the amount
of complexity in sea urchin morphology.

\section{DISCUSSION}

\label{sec:discusion}

Traditionally, radial symmetry has been associated with regularity
while bilateral symmetry with irregularity. In this work we propose
the eutacticity as a measure of the regularity of a biological form
which is independent of the of the radial or bilateral condition. With
this hypothesis, we have shown that regularity has dominance over
irregularity in sea urchins evolution; despite that variability
increases over time, statistically sea urchins show a high degree of
regularity. This regularity is nearly perfect in the most primitive
groups, belonging to the paleozoic era, which were almost totally
radial. A slight decreasing of regularity is observed in
post-paleozoic sea urchins, with the notably exception of the
Holasteroida order which seems to constitute a critical evolutive
event in sea urchins evolution.

\textbf{Acknowledgments}. This work was inspired by Manuel Torres,
whose death will not dismiss the memory of his achievements and
creativity; he will be greatly missed by us. Useful suggestions from
M.E. Alvarez-Buylla, G. Cocho, A. Laguarda and F. Sol\'{\i}s are
gratefully acknowledged. This work was financially supported by the
Mexican CONACyT (Grant No. 50368) and the Spanish MCYT (Grant
No. FIS2004-03237).














\end{document}